\documentclass[apj,twocolumn]{openjournal}
\usepackage[utf8]{inputenc}
\usepackage{amsmath}
\usepackage{hyperref}
\usepackage{yhmath}

\newcommand{\md}{\mathrm{d}}

\usepackage{color}
\usepackage[T1]{fontenc}  
\usepackage{aas_macros}

\usepackage{comment}
\usepackage{cancel}
\usepackage{natbib}
\setcitestyle{numbers}

\begin{document}

\rightline{\scriptsize RBI-ThPhys-2025-33}
\title{Rapid cosmological inference with the two-loop matter power spectrum  \vspace{-1cm} }
\author{Thomas Bakx$^{\dagger,1}$}
\author{Henrique Rubira$^{\dagger\dagger,2,3,4}$}
\author{Nora Elisa Chisari$^{1,5}$}
\author{Zvonimir Vlah$^{6,7,8}$}
\email{$\dagger$ t.j.m.bakx@uu.nl}
 \email{$\dagger\dagger$ henrique.rubira@lmu.de}
\affiliation{$^1$
Institute for Theoretical Physics, 
Utrecht University, Princetonplein 5, 3584 CC, Utrecht,
The Netherlands,
}%
\affiliation{$^2$University Observatory, Faculty of Physics, Ludwig-Maximilians-Universit\"at, Scheinerstr. 1, D-81679 München, Germany}
\affiliation{$^3$Kavli Institute for Cosmology Cambridge, Madingley Road, Cambridge CB3 0HA, UK}
\affiliation{$^4$Centre for Theoretical Cosmology, Department of Applied Mathematics and Theoretical Physics
University of Cambridge, Wilberforce Road, Cambridge, CB3 0WA, UK}
\affiliation{$^5$Leiden Observatory, 
Leiden University, 
Niels Bohrweg 2, NL-2333 CA Leiden, 
The Netherlands.}
\affiliation{$^6$Division of Theoretical Physics, Ruđer Bo\v{s}kovi\'c Institute, 10000 Zagreb, Croatia,}%
\affiliation{$^7$Kavli Institute for Cosmology, University of Cambridge, Cambridge CB3 0HA, UK}%
\affiliation{$^8$Department of Applied Mathematics and Theoretical Physics, University of Cambridge, Cambridge CB3 0WA, UK.}

\begin{abstract}
We compute the two-loop effective field theory (EFT) power spectrum of dark matter density fluctuations in $\Lambda$CDM using the recently proposed \texttt{COBRA} method \cite{cobra}. With \texttt{COBRA}, we are able to evaluate the two-loop matter power spectrum in $\sim 1$ millisecond at $ \sim 0.1 \%$ precision on one CPU for arbitrary redshifts and on scales where perturbation theory applies. As an application, we use the nonlinear matter power spectrum from the Dark Sky simulation to assess the performance of the two-loop EFT power spectrum compared to the one-loop EFT power spectrum at $z=0$. 
We find that, for volumes typical for Stage IV galaxy surveys, $V = 25 \,(\text{Gpc}/h)^3$, the two-loop EFT can provide unbiased cosmological constraints on $\Omega_m,H_0$ and $A_s$ using scales up to $k_\text{max}=0.26\, h/\text{Mpc}$, thereby outperforming the constraints from the one-loop EFT ($k_\text{max}=0.11\, h/\text{Mpc}$). The Figure of Merit on these three parameters increases by a factor $\sim 2.6$ and the one-dimensional marginalized constraints improve by $\sim35\%$ for $\Omega_m$, $\sim20\%$ for $H_0$ and $\sim 15\%$ for $A_s$. 
     
\end{abstract}

\maketitle

\section{Introduction}
The large-scale structure (LSS) of the Universe has become a central probe of cosmological information \cite{BOSS:2016wmc,Abbetal,eBOSS:2019dcv, DESI:2016fyo,Amendola:2012ys,Ivezic:2008fe,SPHEREx:2014bgr,PFSTeam:2012fqu}. To extract the maximal amount of information out of galaxy surveys, modeling the nonlinear process of structure formation is vital. The effective field theory of LSS (EFT of LSS) \cite{Baumann:2010tm, Carrasco:2012cv, Carrasco:2013mua, Konstandin:2019bay,Angulo:2015eqa,Vlah:2015, Baldauf:2021zlt,BOSS:2013eso,DAmico:2019fhj,Ivanov:2019hqk,Colas:2019ret,Chen:2021} builds on top of the nonlinear couplings and loop-integral computations of the standard perturbation theory (SPT) approach \cite{Bernardeau:2001qr}, at the same time marginalizing over the unknown UV physics via counterterms. These approaches have been shown to yield accurate predictions for the power spectrum, bispectrum and trispectrum \cite{Foreman:2015lca,steele_bisp,steele_trisp,taule_bisp} of the dark matter density contrast into the mildly nonlinear regime. Such findings strongly suggest that there is cosmological information to be gained from modelling the matter distribution on scales that are amenable to analytic techniques.

Crucially, rendering such an approach computationally feasible for cosmological inference is by no means trivial, as it requires recalculating high-dimensional integrals (i.e.~loop corrections to the linear theory prediction) at each new set of cosmological parameters. To this end, analytic methods based on Fast Fourier Transforms (FFT) of the linear power spectrum $P_L(k)$ have been developed in \cite{fft_pt, McEwen:2016fjn, vlah_twoloop, Simonovic:2017mhp} and extensively applied to compute power spectra in next-to-leading order (one-loop). Alternative basis including massive propagators were considered by \cite{Anastasiou:2022udy}. However, the calculation of higher-order corrections is either challenging or outright intractable within these techniques. This situation is therefore unsatisfactory: most pressingly, studies from \cite{Baldauf:2021zlt, Garny:2022fsh, Taule:2023izt} offer evidence that two-loop corrections to the power spectrum can significantly increase the range of scales modeled by perturbation theory, but a practically efficient approach to compute such corrections was not known. Numerical evaluation of the 5-dimensional two-loop integrals takes typically in between $\sim 10^1$ and $10^2$ seconds for a single $k$ mode (for $10^{-3}$ relative precision using the code described in \cite{Blas:2013aba, Konstandin:2019bay} and the SUAVE algorithm based on \cite{Lepage:1977sw, Lepage:2020tgj} and implemented in the \texttt{CUBA} library \cite{Hahn:2004fe}), becoming then unfeasible if one wants to repeatedly compute them within a Markov chain Monte Carlo (MCMC) setup to perform cosmological parameter inference.

Recently, in \cite{cobra} (hereafter Paper I) a new, unified approach to this problem  dubbed Cosmology with Optimally factorized Bases for Rapid Approximation (\texttt{COBRA}) was proposed, providing a clear and pragmatic path towards calculating arbitrary corrections to any N-point function. \texttt{COBRA} is based on a decomposition of the linear power spectrum in terms of a representative basis. The loop integrals of EFT of LSS can then be (pre)calculated for this basis, such that when exploring different cosmologies within an MCMC one only has to project the given cosmology onto this integral basis.  
Importantly, the proposed technique is optimal in terms of efficiency and memory requirement and does not rely on analytic expressions for perturbation theory integrals, unlike previous approaches \cite{Simonovic:2017mhp}. 

A first case study comprising the one-loop power spectrum of galaxies yielded encouraging results, but as already pointed out in Paper I the efficacy of \texttt{COBRA} is most prominent in more complicated scenarios. In this paper, we apply \texttt{COBRA} to the two-loop power spectrum of dark matter fluctuations. We show that \texttt{COBRA} is able to reproduce the direct two-loop result with permille precision for a broad range of cosmologies. Finally, we run an MCMC with the two-loop matter power spectra, indicating its relative gains with respect to the one-loop scenario (for earlier results in this direction, see \cite{cataneo_expl,osato_2loop1,osato_2loop2}).

The structure of the paper is the following. In Section \ref{sec:cobra} we briefly summarize the \texttt{COBRA} construction from Paper I. In Section \ref{sec:cobra2L}, we review the two-loop EFT of LSS theory, including the so-called IR resummation and compare the \texttt{COBRA} two-loop calculation to the direct two-loop result. We then explicitly apply it to the EFT prediction for the clustering of dark matter at two loops and compare it to the nonlinear matter power spectrum measured from the Dark Sky simulation \cite{skillman} in \ref{sec:simulation}. We conclude in Section \ref{sec:disc}. Some technical details regarding our numerical implementation of the two-loop power spectrum are relegated to Appendix \ref{sec:tensors}, we discuss the IR-safety of the loop integrals in \ref{app:IRsafe} and present the full MCMC posteriors in \ref{app:fullMCMC}.

\section{COBRA}\label{sec:cobra}

The main idea of \texttt{COBRA} is to achieve an optimal decomposition of the linear matter power spectrum into a small set of basis functions of the form 
\begin{eqnarray}\label{eq:decomp}
    P_L^\Theta(k) = \sum_{i=1}^{N_b}w_i(\Theta)v_i(k)\,,
\end{eqnarray}
where the number of terms $N_b$ is as small as possible to guarantee a required precision in the loop computations. Here $v_i$ are functions that depend only on the wavenumber $k$, called \textit{scale functions}, and the coefficients $w_i(\Theta)$ depend only on cosmological parameters, collectively labeled $\Theta$. Obtaining such a decomposition over a given range of cosmological parameters is relatively cheap and can be straightforwardly implemented via a singular value decomposition (SVD). We consider in this work $\Lambda$CDM cosmologies with priors described in Table~\ref{tab:priors}. For this prior range, it is possible to achieve an accuracy of below $\sim 10^{-3} \, (10^{-4})$ with $N_b = 9 \, (12)$ basis functions in the {\it full}, i.e. containing wiggles, linear matter power spectrum. For the {\it no-wiggle} spectra used in the IR-resummation procedure [see Section~\ref{sec:IR} for definition], we consider fewer elements in the basis since convergence is typically faster than for the full spectra. The priors over the cosmological parameters considered in this work, as described in Table~\ref{tab:priors}, are exactly the same as Paper I, but the $v_i$ functions are slightly different due to the larger range in wavenumber $k$.

\begin{table}[t]
    \centering
    \begin{tabular}{|c|c|c|}
    \hline
         & \multicolumn{2}{c|}{Default} \\
        \hline
        $\Theta $& Emulation range for $w_i(\Theta)$ & Grid size for SVD  \\
        \hline\hline
        $\omega_c$ & [0.095,0.145] & 30 \\ 
        $\omega_b$ & [0.0202,0.0238] & 15 \\
        $n_s$ & [0.91,1.01] & 15 \\
        $10^9 A_s$ & - & $10^9 A_s^* = 2$  \\
        $h$ & [0.55,0.8] & $h^* = 0.7$ \\
        $z$ & - & $z^* = 0$ \\
        \hline
    \end{tabular}
    \caption{Ranges and template grids for the $\Lambda$CDM cosmological parameters used to obtain the basis functions $v_i$ for \texttt{COBRA}. If a parameter is held fixed for the templates, its fiducial value is indicated. The templates are always linearly spaced across the indicated range. For $A_s$, the power spectrum scaling is trivial. 
    When $\omega_b$ and $\omega_c$ are held fixed, $h$ also only rescales the amplitude of the power spectrum and hence is held fixed when constructing the templates. 
    We emulate $w_i(\Theta)$ as a function of $\omega_c,\omega_b, n_s $ and $h$.  }
    \label{tab:priors}
\end{table}

In order to obtain the decomposition from Eq.~\eqref{eq:decomp} for a set of cosmology-independent basis functions, it is important to perform the decomposition in cosmology-independent units; to this end we define $h^*=0.7$ and we choose the range $0.0008 \,h^*/\text{Mpc} < k < 20\,h^*/\text{Mpc}$. 
We use \texttt{CAMB} \cite{camb} (v1.5.2) as our Boltzmann solver. Once the set of basis functions $v_i$ is obtained via the SVD decomposition, the weights can either be computed via projection onto the linear power spectrum from a Boltzmann solver, or emulated through an alternative method. Here we employ the latter approach, utilizing the RBF interpolation detailed in Paper I (see \cite{fasshauer} for the original description of the method). This scheme is fast and emulates the weights with negligible accuracy loss (see e.g.~Fig. 4 of \cite{cobra}). 

Using the linear decomposition from Eq. \eqref{eq:decomp}, loop integrals can be recast as small multilinear contractions involving precomputed tensors $\mathcal{S}_{ij\dots}$ and the weights $w_i(\Theta)$, where the tensors do not depend on cosmological parameters. Each index of the tensor runs over the $N_b$ elements of the basis $v_i(k)$. As an example, a one-loop power spectrum can be written as
\begin{eqnarray}\label{eq:1loop1}
    P^\Theta_{\text{1-loop}}(k) &=& \text{st.}(k) + \mathcal{S}^{l}_{i}(k)w_i(\Theta) \nonumber\\
    &+& \mathcal{S}^{q}_{ij}(k)w_i(\Theta)w_j(\Theta)\,,
\end{eqnarray}
where the sum over indices is implicit, $\mathcal{S}^{l}_i$ and $\mathcal{S}^{q}_{ij}$ are respectively linear and quadratic in the power spectra and the first term is independent of $P_L$, i.e.~a stochastic component \cite{Rubira:2024tea}. The two-loop power spectrum involves cubic integrals of $P_L(k)$, demanding then an extra rank-3 tensor $\mathcal{S}^{c}_{ijk}$ so that
\begin{eqnarray}\label{eq:2loop1}
    P^\Theta_{\text{2-loop}}(k) &=& \text{st.}(k) + \mathcal{S}^{l}_{i}(k)w_i(\Theta) \nonumber \\ 
    &+& \mathcal{S}^{q}_{ij}(k)w_i(\Theta) w_j(\Theta)  \nonumber \\ 
    &+& \mathcal{S}^{c}_{ijk}(k)w_i(\Theta) w_j(\Theta) w_k(\Theta)\, . 
\end{eqnarray}
It is important to notice that the considered number of elements in the basis is 
not necessarily the same for all $\mathcal{S}_{ij\dots}$ tensors, and this can be calibrated according to the precision required for each term. See Appendix~\ref{sec:tensors} for a broader discussion on how the tensors are calculated. The most time consuming part of \texttt{COBRA} is the precomputation of those tensors, since the projection onto the basis that happens during parameter inference is numerically cheap. Considering fewer elements on the basis for the tensors $\mathcal{S}_{ij\dots}$ also reduces computational cost, since one has to precompute fewer loop integrals. We discuss the number of elements in the basis considered for each loop in Section~\ref{sec:results_fast2L}, but we mention in advance that for this work we precomputed $165$ ($56$) two-loop integrals for the full (no-wiggle) spectra in every $k$-bin to get $\sim 0.1\%$ precision on the total matter power spectrum. 

\section{Two-loop Power Spectrum with COBRA}\label{sec:cobra2L}

In this Section, we briefly review the EFT of LSS formalism in Section \ref{sec:eft2L}, then comment on the IR-resummation in Section \ref{sec:IR}. We discuss how fast and precise the two-loop computation is with \texttt{COBRA} in Section \ref{sec:results_fast2L}. We briefly comment on the UV sensitivity of the loop integrals with \texttt{COBRA} in Section~\ref{sec:UVsens}.

\subsection{The two-loop EFT} \label{sec:eft2L}

In this work, we focus on the matter nonlinear power spectrum. The description of matter tracers would involve including bias operators (see \cite{Desjacques:2016bnm} for a review), which is beyond the scope of this work.
The one-loop matter power spectrum in the EFT of LSS contains only one free parameter; it is given by \cite{Baumann:2010tm, Carrasco:2012cv} 
\begin{eqnarray}\label{eq:oneloop}
  P_{\text{EFT,1l}} = P_L + P_{\text{SPT,1l}} - 2(2\pi)c_{\text{s},1}^2 \frac{k^2}{k_\text{nl}^2}P_L  \,,
\end{eqnarray}
where $P_{\text{SPT,1l}}$ is the one-loop SPT spectrum, i.e.
\begin{eqnarray}
    P_{\text{SPT,1l}} = P_{22} + 2 P_{13}\,,
\end{eqnarray}
and $c_{\text{s},1}^2$ is the sound velocity counterterm associated with the $\nabla^2 \delta$ operator. To render the counterterms dimensionsless, we normalize the higher-derivative counterterm by the nonlinear scale, which we take to be $k_\text{nl} = 1\,h/$Mpc for simplicity when performing the fits. The two-loop power spectrum in the EFT of LSS framework is given by  \cite{Carrasco:2013mua,Foreman:2015lca,Konstandin:2019bay}
\begin{eqnarray}\label{eq:twoloops}
    &&P_{\text{EFT,2l}}  =P_L  + P_{\text{SPT,1l}} + P_{\text{SPT,2l}} - 2(2\pi)c_{\text{s}}^2 \frac{k^2}{k_\text{nl}^2}P_L \nonumber \\
    &-& 2(2\pi)c_{\text{s},1}^2 \frac{k^2} {k_\text{nl}^2}P_{\text{SPT,1l}} - 2(2\pi)c_{\text{quad}} \frac{k^2} {k_\text{nl}^2}P_{\text{quad}} \nonumber\\
    &+&(2\pi)^2 [(c_{\text{s},1}^2)^2-2c_4]\frac{k^4} {k_\text{nl}^4}P_L + (2\pi)^2 c_\epsilon \frac{k^4} {k_\text{nl}^4} \frac{1}{k_\text{nl}^3}\,,
\end{eqnarray}
where $P_{\text{SPT,2l}}$ is the complete two-loop SPT spectrum, i.e. \cite{Simonovic:2017mhp} 
\begin{eqnarray}\label{eq:twoloopspt}
    P_{\text{SPT,2l}} = P_{33}^I + P_{33}^{II} + 2P_{24} + 2P_{15} \,,
\end{eqnarray}
and $c_\text{s}^2 = c_{\text{s},1}^2+c_{\text{s},2}^2$. The two-loop EFT Eq.~\eqref{eq:twoloops} has counterterms $c_{\text{s},1}^2$ and $c_{\text{s},2}^2$ linked to the $\nabla^2\delta$ operator and $c_{\rm quad}$ to the quadratic $\nabla^2 \delta^2$ term, i.e.
\begin{eqnarray}
    P_\text{quad} = \int_\mathbf{q}\, F_2(\mathbf{q},\mathbf{k-q})P_L(q)P_L(|\mathbf{k-q}|).
\end{eqnarray}
Finally, $c_4$ corresponds to the quartic-derivative $\nabla^4\delta$ and  $c_\epsilon$ to the two-loop stochastic correction, totaling five free parameters. In what follows, we set $c_\epsilon = 0$ in line with the findings of \cite{Foreman:2015lca}. Hence, we only consider the four parameters 
\begin{eqnarray}\label{eq:renlist}
    \{c_{\text{s},1}^2, c_{\text{s},2}^2,  c_{4}, c_{\text{quad}}\}\,,
\end{eqnarray}
for the two-loop EFT.

\subsection{IR resummation} \label{sec:IR}

In complete analogy to Paper I, we implement the resummation of the large-scale displacements (IR resummation) using the `wiggle-no-wiggle split' \cite{baldauf_bao, vlah_osc, Blas:2016sfa} to separate the broadband power from the BAO feature, where the no-wiggle linear power spectrum is, following \cite{vlah_osc}, given by 
\begin{eqnarray}\label{eq:dewig}
    P^\Theta_{\text{nw}}(k) &=& P^*_\text{EH}(k)\frac{1}{\lambda(\log k)\sqrt{2\pi}}\int \md \log{q}\frac{P^\Theta_L(10^{\log{q}})}{P^*_\text{EH}(10^{\log{q}})} \nonumber \\ 
    &\times & \exp{\bigg(-\frac{1}{2\lambda(\log k)^2}(\log{q}-\log{k})^2\bigg)}\,,
\end{eqnarray}
with $P^*_\text{EH}(k)$ the Eisenstein-Hu power spectrum \cite{Eisenstein:1997jh} and $\lambda(k)$ is a scale-dependent smoothing scale.\footnote{Explicitly, the filter takes the form 
\begin{eqnarray}
    \lambda(\log k) = \alpha_1 + \alpha_2\, \frac{\exp[-\alpha_3(\log k -\alpha_4)^2]}{1+\exp[\alpha_5(\log k + \alpha_6)]}\,,
\end{eqnarray}
with coefficients $\alpha_1 = 0.005,\,\alpha_2=0.45,\,\alpha_3=0.9,\,\alpha_4=-0.15,\,\alpha_5=8,\,\alpha_6 = 0.5$.
We experimented with other choices and found only minor impact on the quality of the fit.}
The filtering operation $\mathcal{F}$ of smoothing the power spectrum wiggles is linear and independent of cosmology and therefore 
\begin{eqnarray}\label{eq:nowigdec}
    P_{\text{nw}}^\Theta(k) &=& \mathcal{F}[w_i(\Theta)v_i](k) = w_i(\Theta) \mathcal{F}[v_i](k)\nonumber \\
    &=& w_i(\Theta) v_{i}^{\text{nw}}(k),
\end{eqnarray}
and correspondingly 
\begin{eqnarray}\label{eq:wigglydec}
    P_{\text{w}}^\Theta(k) &=& w_i(\Theta) \left[v_i(k) - \mathcal{F}[v_i](k)\right] \nonumber \\ 
    &=& w_i(\Theta) v_{i}^{\text{w}}(k).
\end{eqnarray}

The damping factor that is needed for the IR-resummation is defined as
\begin{eqnarray}
    \Sigma^2(\Theta) \equiv \int_0^{k_\text{o}}\frac{\md q}{6\pi^2}\left[1-j_0(qr_\text{o}) + 2j_2(qr_\text{o})\right]P_\text{nw}(q),
\end{eqnarray}
and we used $r_\text{o} = 105 \text{ Mpc}/h^*$ and $k_\text{o} = 0.2\, h^*/\text{Mpc}$. 
In addition to this IR resummation scheme, which relies on the somewhat arbitrary choice of a wiggle–no-wiggle split, there exist several direct IR resummation approaches primarily based on Lagrangian perturbation theory (see, e.g., \cite{Senatore:2014via, Vlah:2015,Schmittfull:2018}). As a cross-check of our results, we also implemented the scheme used in \cite{Foreman:2015lca} to verify that any potential differences in our results do not stem from the choice of IR resummation and by its possible (albeit small) mixing of scales in the UV regime. A comparison of the two resummation schemes reveals no significant differences between the results.

At one-loop, IR resummation (using the wiggle-no-wiggle split) amounts to replacing the linear power spectrum by 
\begin{eqnarray}
    P_{L} \rightarrow P_{\text{nw}} + e^{-k^2 \Sigma^2}(1+k^2 \Sigma^2) P_{\text{w}}.
\end{eqnarray}
The one-loop correction is replaced by
\begin{eqnarray}
    P_{\text{SPT,1l}} \rightarrow P_{\text{nw},\text{1l}}+ e^{-k^2 \Sigma^2} P_{\text{w},\text{1l}}\,,
\end{eqnarray}
and for the one-loop counterterm we use  
\begin{eqnarray}
    \frac{k^2}{k_\text{nl}^2} P_{L} \rightarrow \frac{k^2}{k_\text{nl}^2}\left[P_{\text{nw}} + e^{-k^2 \Sigma^2}P_{\text{w}}\right].
\end{eqnarray}
Here and in the below, `wiggly' variants for power spectra are defined as $P_{\text{w},X} = P_{\text{full},X}-P_{\text{nw},X}$. At two loops, we implement IR resummation by replacing 
\begin{eqnarray}
\begin{aligned}
    P_{L} &\rightarrow P_{\text{nw}} + e^{-k^2 \Sigma^2}\left[1+k^2 \Sigma^2 + \frac{1}{2}(k^2 \Sigma^2)^2\right] P_{\text{w}}\,,
\end{aligned}
\end{eqnarray}
in the linear theory part of the prediction, while the one- and two-loop SPT part are replaced by 
\begin{eqnarray}
\begin{aligned}
     P_{\text{SPT,1l}} &\rightarrow P_{\text{nw},\text{1l}}+ e^{-k^2 \Sigma^2}(1+k^2 \Sigma^2) P_{\text{w},\text{1l}} \,,  \\
     P_{\text{SPT,2l}} &\rightarrow P_{\text{nw},\text{2l}}+ e^{-k^2 \Sigma^2} P_{\text{w},\text{2l}}\,
\end{aligned}
\end{eqnarray}
and the `quad' part by 
\begin{eqnarray}
    \frac{k^2}{k_\text{nl}^2}P_{\text{quad}} \rightarrow \frac{k^2}{k_\text{nl}^2}\left(P_{\text{nw},\text{quad}}+ e^{-k^2 \Sigma^2} P_{\text{w},\text{quad}}\right).
\end{eqnarray}
The two other counterterms become 
\begin{eqnarray}
    \begin{aligned}
        \frac{k^2}{k_\text{nl}^2} P_{L} &\rightarrow \frac{k^2}{k_\text{nl}^2} \left[P_{\text{nw}} + e^{-k^2 \Sigma^2}(1+k^2 \Sigma^2) P_{\text{w}}\right] \,, \\
        \frac{k^4}{k_\text{nl}^4} P_{L} &\rightarrow \frac{k^4}{k_\text{nl}^4}(P_{\text{nw}} + e^{-k^2 \Sigma^2}P_{\text{w}}).
    \end{aligned}
\end{eqnarray}
Moreover, we also implemented the one and two-loop integrals using the IR-safe integrals \cite{Carrasco:2013mua,Blas:2013aba}, which we discuss in details in Appendix~\ref{app:IRsafe}.

\subsection{COBRA two-loop results}\label{sec:results_fast2L}

In this section we discuss one of the main results of this paper, that is the fast computation of the two-loop EFT spectra with \texttt{COBRA}. 
The complete two-loop power spectrum requires the calculation of rank-two (one-loop) $\mathcal{S}^{q}_{ij}$ and rank-three (two-loop) $\mathcal{S}^{c}_{ijk}$ tensors by inserting combinations of the basis functions $v_j(k)$ into the loop integrals, according to Eqs.~\eqref{eq:1loop1} and \eqref{eq:2loop1}, respectively. We set the required numerical relative precision of the Monte Carlo integration to $10^{-3}$ for all integral evaluations as well as a maximum number of $8\times 10^7$ Monte Carlo evaluations. We anticipate that permille precision in the loops is sufficient to match the expectations of ongoing galaxy surveys.
We use the two-loop codes of \cite{Blas:2013aba, Konstandin:2019bay} and \cite{vlah_osc, Fasiello:2022} together with the \texttt{CUBA} numerical integration library \cite{Hahn:2004fe} and the SUAVE algorithm based on \cite{Lepage:1977sw, Lepage:2020tgj}. 

For each loop term, both the full and no-wiggle tensors need to be computed (that is, with $v_j(k)$ for the `full' and $v_j^{\rm nw}(k)$ for the `no-wiggle' term as inputs). We use different output wavenumber grids for both cases, since the `full' term always includes wiggly features that need to be more densely sampled. For the full terms, we use $50$ equally log-spaced wavenumbers between $0.01\, h^*/$Mpc and $1\, h^*/$Mpc with $15$ extra points (total $65$ points) between $0.2\, h^*/$Mpc and $0.6\, h^*/$Mpc to increase the resolution in this region. For the no-wiggle part, which is much smoother, we used $30$ equally log-spaced wavenumbers between $0.01\,h^*/$Mpc and $1\,h^*/$Mpc.\footnote{To understand whether we have used a sufficient number of wavenumber bins so as to not suffer from residual interpolation errors, we also computed the two-loop power spectrum at one fixed cosmology in $200$ equally log-spaced bins and found negligible differences.}

\begin{figure}
    \centering
    \includegraphics[width=0.95\linewidth]{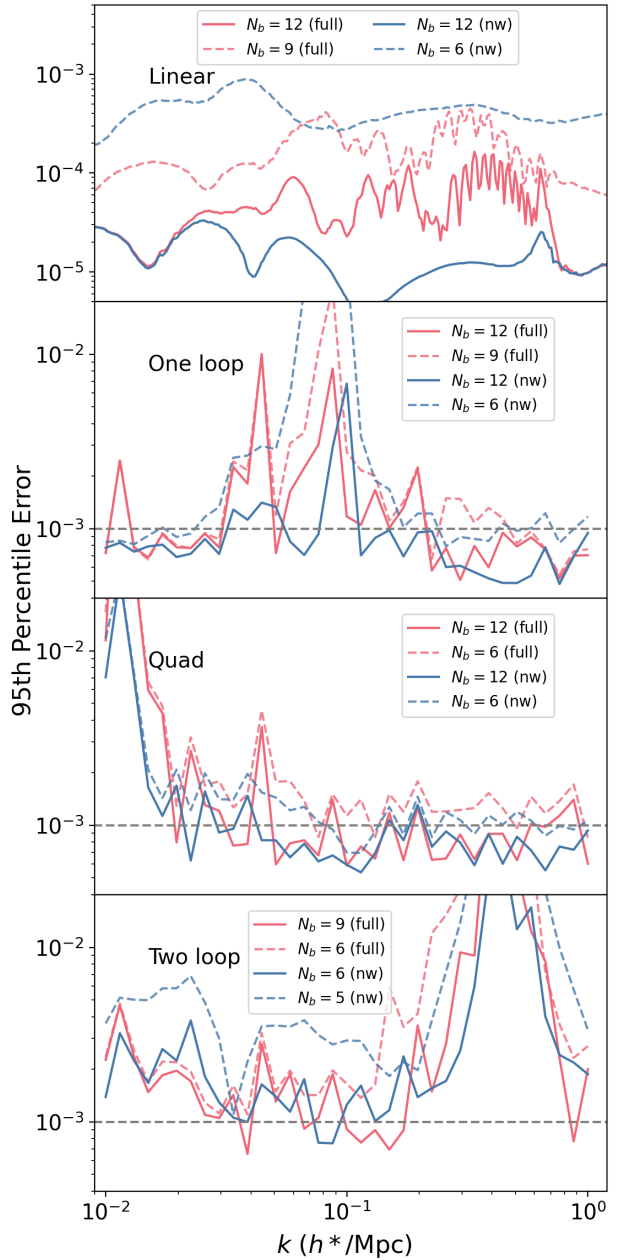}
    \caption{Error on individual contributions to the two-loop EFT power spectrum using the \texttt{COBRA} expansion relative to the non-expanded numerical result. They are calculated as the 95th percentiles of a total of 200 random test cosmologies within the \texttt{COBRA} prior range. We display the full and no-wiggle spectra in red and blue, respectively. The linear spectra, the one-loop SPT term, the quadratic counterterm and the two-loop contributions are shown in different rows. In each subplot, we show two separate choices of $N_b$ in different line styles, where the solid lines correspond to Eqs. \eqref{eq:nbnw} and \eqref{eq:nbtot}.    
    The spikes in the second and fourth panels are largely due to the integrals taking values close to zero, which are suppressed when considering the full spectra (see Fig. \ref{fig:total_result}). The `quad' term (third panel) scales as $k^4$ on large scales and thus presents larger relative errors at low $k$. The black dashed lines indicate the $10^{-3}$ required precision of the numerical integrals.}
    \label{fig:terms_result}
\end{figure}

We compare the two-loop power spectrum across a set of 200 randomly selected test cosmologies at $A_s^* = 2.1 \times 10^{-9}$ and $z^* = 0$ within the specified parameter ranges of Table~\ref{tab:priors} for $\{\omega_b,\omega_c, n_s, h\}$, so that we can explicitly assess the performance of \texttt{COBRA}. For these test spectra, we use 35 equally log-spaced bins in the interval $[0.01,1]\,h^*/$Mpc. The 95th percentile of the error in each of the two-loop terms is shown in Fig.~\ref{fig:terms_result} for the no-wiggle and full spectra. The error is calculated relative to the full numerical result without using the \texttt{COBRA} basis. The horizontal dashed line indicates the $10^{-3}$ precision required by the \texttt{CUBA} numerical integrator (see the beginning of this Subsection), setting then a lower bound for the ratio. We show errors for different values of $N_b$ used in the computation of the \texttt{COBRA} tensors.

For the one-loop and `quad' integrals, we find excellent agreement between \texttt{COBRA} and the non-projected integrals throughout most of the $k$ range considered, with most of the points being close to the $10^{-3}$ lower numerical bound. That indicates that further increasing the number of elements in the basis has little impact compared to increasing the relative precision of the loop integrals. The error in the two-loop result averages around $2\times10^{-3}$ on most scales, still potentially benefiting from increasing $N_b$ further. We stress however that this is not needed in the comparison in Section \ref{sec:simulation}. Moreover, note that one typically needs less elements for the no-wiggle spectra compared to the full spectra to achieve the same precision, since it contains fewer features. 
We also find spikes in regions where the loop spectra typically cross zero, which, since it varies for different cosmologies it appears as a broad region in $k$. Those spikes are also reduced by increasing the number of elements in the basis (see e.g. the comparison between solid and dashed lines around $k \approx 0.4\,h/\text{Mpc}$ for the two-loop contributions). The lack of precision around the zero-crossing of the loops is suppressed when considering the full spectra, as discussed below.

Throughout this work, we use
\begin{equation} \label{eq:nbnw}
    N_b^{\rm nw} = \{12,12,12,6\}\,,
\end{equation} 
for linear, one-loop, `quad' counterterm and two-loop no-wiggle contributions and 
\begin{equation} \label{eq:nbtot}
    N_b^{\rm full} = \{12,12,12,9\}\,,
\end{equation} for the full contributions, respectively. As shown in Fig.~\ref{fig:terms_result}, sticking to this choice of the number of elements in the basis, we generally find sub-percent precision for the individual terms for the range of $k$ considered in this work. As we will see below, these numbers are sufficient to obtain sub-permille accuracy on the two-loop power spectrum. 

For the construction of the two-loop tensors, adopting those numbers of elements in the tensorial basis leads to computing 165 and 56 integrals respectively for the full and no-wiggle spectra, per $k$ point (see App.~\ref{sec:tensors}). Therefore, we only had to store around 200 float numbers per mode, which leads to no memory issues.
In terms of computation power, as commented before, \texttt{COBRA} precomputes the loop spectra in the SVD basis, such that further evaluation for a given cosmology only demands projecting the linear power spectra to obtain the $w_i$ coefficients, which takes $\sim 1$ ms on one CPU in Python 3. The precomputation of the two-loop spectra (which corresponds to the main time bottleneck) in $\mathcal{S}^{c}_{ijk}$ takes typically less than $10^2$ seconds per mode (for $10^{-3}$ relative precision) in a single core. Overall, the \texttt{COBRA} implementation of the two-loop integrals is efficient in both memory usage and computational performance.

\begin{figure}
    \centering
    \includegraphics[width=\linewidth]{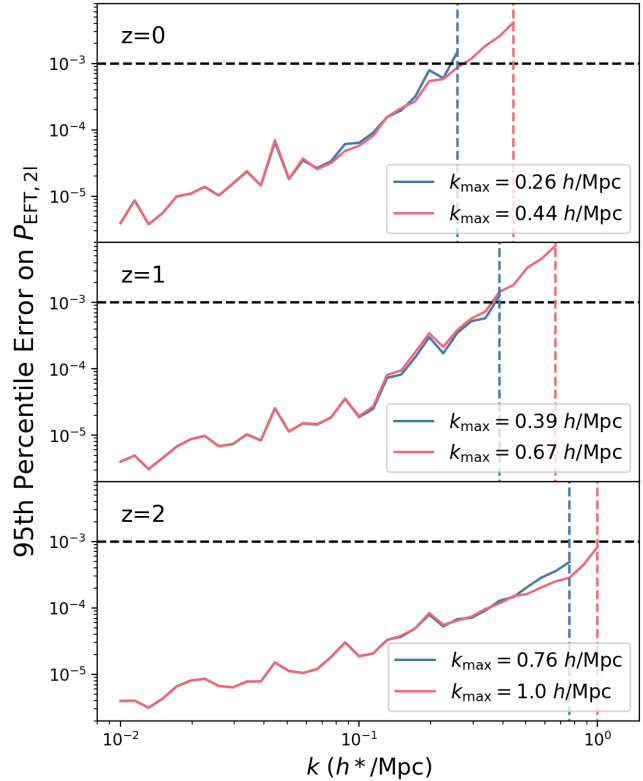}
    \caption{The 95th percentile of errors on the {\it total two-loop spectrum} (cf.~Eq.~\eqref{eq:twoloops}) of the test set of 200 random cosmologies calculated using \texttt{COBRA}, compared to the direct numerical result. We obtain the counterterms by fitting the two-loop spectra to the \texttt{HaloFit} at three different redshifts, $z=0,1,2$ using two different values of $k_\text{max}$ at all redshifts (vertical dashed lines). We include the IR resummation and take $N_b^\text{nw} = \{12,12,12,6\}$ basis functions for no-wiggle contributions and $N_b^\text{full} = \{12,12,12,9\}$ for the full contributions, respectively, as in Eqs. \eqref{eq:nbnw}
    and \eqref{eq:nbtot}.}
    \label{fig:total_result}
\end{figure}

To assess the precision of the total \texttt{COBRA} EFT result, summing over all the terms, we also need to choose appropriate values for the counterterms. We use the nonlinear power spectrum template obtained from the \texttt{HaloFit} \cite{Takahashi:2012em} as implemented in \texttt{CCL} \cite{LSSTDarkEnergyScience:2018yem} (evaluated at a fiducial \textit{Planck}-like cosmology) for fitting the counterterms at three different redshifts $z=0,1,2$. We use two different scale cuts at each redshift. The best-fit counterterms obtained from fitting Eq. \eqref{eq:twoloops} to \texttt{HaloFit} are then used to compute the total two-loop power spectrum either with \texttt{COBRA} or directly at the test cosmology. 
We show the 95th percentile of resulting errors in the total power spectrum from Eq. \eqref{eq:twoloops} (including IR resummation) in Fig. \ref{fig:total_result}, calculated for the same 200 test sample cosmology sample described above. The two choices of $k_{\rm max}$ are meant to roughly bracket the nonlinear scale at that redshift, see e.g. \cite{Foreman:2015lca}. 
The errors on the total power spectrum are generally sub-permille, and increase towards larger $k$ where the two-loop contributions (which dominate the total error) become more important. Similar considerations also explain why at a given wavenumber $k$, the performance improves when going to higher redshift, since the two-loop contribution is suppressed by powers of the growth factor. The larger errors at higher $k_\text{max}$ can be further ameliorated by increasing the precision of the numerical integration, or the number of basis functions used for the two-loop calculation.

We therefore conclude that \texttt{COBRA} achieves remarkable $\sim 10^{-3}$ relative precision at two loops in the considered prior range with the number of elements in the basis given by Eq.~\eqref{eq:nbnw} for the no-wiggle basis and Eq.~\eqref{eq:nbtot} for the full spectra basis. This is the first time, to the knowledge of the authors, that such a fast and accurate calculation of the two-loop matter power spectra is provided.

\subsection{UV sensitivity} \label{sec:UVsens}
\begin{figure}
    \centering
    \includegraphics[width=\linewidth]{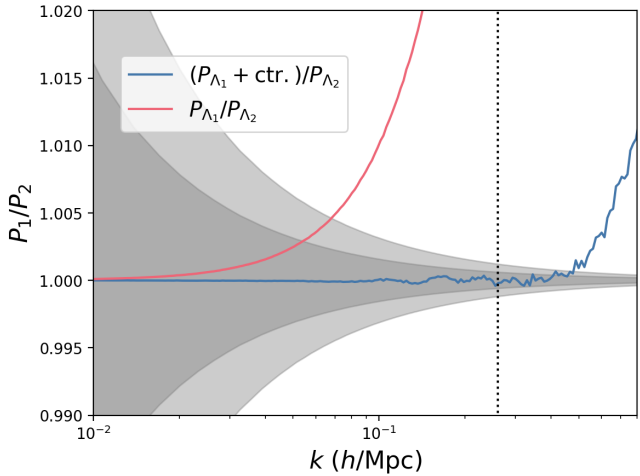}
    \caption{Illustration of the UV-sensitivity of the two-loop SPT prediction [cf.~Eq.~\eqref{eq:twoloopspt}] and how it can be absorbed by counterterms. We compute the two-loop matter power spectrum at two different cutoffs $\Lambda_1 = 2\,h/\text{Mpc}$ and $\Lambda_2 = 4.5\,h/\text{Mpc}$. Their ratio is the red line, which clearly exhibits running beyond the error we can tolerate [gray shades which represent $1\sigma$ and $2\sigma$ errors for a volume of $V_\text{fit} = 25  \,(\text{Gpc}/h)^3$, see Section \ref{sec:simulation}]. The blue line indicates the error on the ratio after adding the counterterms, determined by fitting up to $k=0.4\,h/\text{Mpc}$. The vertical dotted line indicates the maximum scale we consider for the two-loop matter power spectrum in this work (see Section~\ref{sec:results}).} 
    \label{fig:ctr_result}
\end{figure}

To conclude this Section, we also briefly validate our renormalization prescription including the four counterterms from Eq.~\eqref{eq:renlist} as free parameters.\footnote{Other renormalization schemes such as \cite{Foreman:2015lca} (see also \cite{Konstandin:2019bay}) fix the one-loop SPT counterterm at lower $k_{\rm max}$. We adopt here a more conservative approach of letting both $c_s$ terms free.} A central point of the EFT description from Eq.~\eqref{eq:twoloops} is that these several counterterms are meant to absorb the unphysical UV dependence of the SPT loop integrals. To check this we compute the two-loop prediction at two different cutoffs $\Lambda_1 = 2\, h/\text{Mpc}$ and $\Lambda_2 = 4.5\, h/\text{Mpc}$ 
and fit the SPT power spectrum at the latter using the EFT model with the former cutoff. We use the Dark Sky cosmology of Sec.~\ref{sec:DarkSky}.
The agreement between the two is shown in Fig. \ref{fig:ctr_result}. The grey bands indicate the $1$ and $2\sigma$ cosmic variance error bars Eq.~\eqref{eq:cosmicvar} for the volume $V_\text{fit}$ considered in Section~\ref{sec:simulation}. 
We see that after including the counterterms, the two results match within the statistical errors.

\section{Comparison to simulation} \label{sec:simulation}

In this Section we demonstrate the applicability of the \texttt{COBRA} two-loop results comparing that to simulation data, which is described in Sec.~\ref{sec:DarkSky}. We present the metrics used for our analysis in Sec.~\ref{sec:metrics} and the results in Sec.~\ref{sec:results}.

\subsection{The Dark Sky simulation} \label{sec:DarkSky}

For the comparison to data, we used the Dark Sky simulation power spectrum. The \texttt{ds14$\_$a} run of the Dark Sky simulation suite is an extremely large N-body simulation first described in \cite{skillman} containing $10240^3$ particles in a box of side length $L = 8 $ Gpc$/h$. This yields a total volume of $V_\text{full} = 512 \,(\text{Gpc}/h)^3$,  and a fundamental mode of $k_f = 2\pi/L \approx 8\times 10^{-4} \,h/{\rm Mpc}$. We use the available $z=0$ redshift snapshot. As such, the relative cosmic variance error 
\begin{eqnarray}\label{eq:cosmicvar}
    \frac{\sigma_{P(k)}}{P(k)} = \sqrt{\frac{2}{N_k}} = \sqrt{\frac{k_f^3}{2\pi k^2 \Delta k}}\,,
\end{eqnarray}
on the power spectrum is very small, reaching values of $\left[\sigma_{P}/P\right]_{k = 0.25\,h/\text{Mpc}} \approx 6 \times 10^{-4}$ even for our binning choice of $\Delta k = 4 k_f \approx 0.003 \,h/\text{Mpc}$. We neglect the shot-noise contribution, since the number density for matter is very high.

Such volumes and corresponding measurement errors are not within reach of current or next-generation large-scale structure surveys. We therefore take the output power spectrum from the \texttt{ds14$\_$a} run and assign to it a fractional error corresponding to a volume of $V_\text{fit}  = 25  \,(\text{Gpc}/h)^3$, which resembles that of a Stage IV spectroscopic galaxy survey such as the Dark Energy Spectroscopic Instrument (DESI) \cite{DESI:2016fyo}. 

Noticing that the loop contributions are heavily suppressed [scaling as $ k^2 P_L(k)$] on large scales, we extrapolate the loop corrections at low wavenumbers using a power law of the form 
\begin{eqnarray}
    P(k) = P(k_{\rm low})\times \bigg(\frac{k}{k_{\rm low}}\bigg) ^{2+n_s}\,,
\end{eqnarray} 
where $n_s$ is the spectral index at the given cosmology and $k_{\rm low}=0.01\,h^*/\text{Mpc}$ .

\subsection{Likelihood and performance metrics} \label{sec:metrics}

We assume a Gaussian likelihood with a diagonal covariance, where the diagonal elements of the covariance in each $k$-bin are defined via Eq.~\eqref{eq:cosmicvar}, with $P$ therein being the nonlinear spectrum measured from the simulation. 
We expect that the non-Gaussian part of the covariance, that comes from the trispectrum, is relatively suppressed compared to the Gaussian part throughout all the scales considered in this work, (see e.g. \cite{bertolini}).
Hence, the likelihood is given by\footnote{Note that this definition differs from the choice taken in \cite{Konstandin:2019bay}, which was based on the residuals of the fit to the simulation data. In this work we weight the log-likelihood by the power spectrum error bars $\sigma_{P}$, as in Eq.~\eqref{eq:Likelihood}.} 
\begin{eqnarray} \label{eq:Likelihood}
    -2 \log \mathcal{L}(k_\text{max}) = \sum_i^{k_\text{max}}\bigg(\frac{P_{\text{EFT}}(k_i) - P_{\rm sim}(k_i)}{\sigma_{P(k_i)}}\bigg)^2\,,
\end{eqnarray}
where $k_{\rm max}$ is the largest wavenumber considered in our analysis and $k_\text{min} = 0.003\,h/\text{Mpc}$. We discuss the $k_{\rm max}$ choice for both one and two-loop below. We interpolate the \texttt{COBRA} $P_{\text{EFT}}(k_i)$ onto the $k_i$ bins used to measure the simulation power spectrum $P_{\rm sim}$. Once more, we explicitly verified that the interpolation error is negligible by sampling extra $k$ points in a single cosmology. We adhere to the choices from Eqs. \eqref{eq:nbnw} and \eqref{eq:nbtot} for the number of basis functions. We checked that using a lower number for the two-loop terms does not impact the results; we thus conclude that our inference is sufficiently precise.

Following e.g.~\cite{desi_fs, schoneberg_bbn}, we put a conservative BBN prior on the baryon density\footnote{Technically, it is a truncated normal prior cut off at the boundaries of the range indicated in Table \ref{tab:priors}. Note however that these boundaries are $>3\sigma$ away from the mean of the normal distribution, so this has no effect on posteriors.} 
\begin{eqnarray}
    \omega_b \sim \mathcal{N}(0.0221396,0.00055^2),
\end{eqnarray}
i.e., centered at the true value of the simulations and with width equal to that of \cite{schoneberg_bbn}. We additionally put a uniform prior 
\begin{eqnarray}\label{eq:nsprior}
    n_s \sim \mathcal{U}(0.91,1.01)\,,
\end{eqnarray}
which roughly coincides with the normal prior from \cite{desi_fs}. We find that these two parameters are generally not well constrained by the power spectrum, and so we treat them as nuisance parameters to be marginalized over without quoting their posterior means and uncertainties. For the other cosmological parameters, we consider a uniform prior in the range of Table~\ref{tab:priors}. We sample the parameters $\omega_c,h,A_s\times 10^9$ and only transform to $\Omega_m, H_0$ and $A_s\times 10^9$ after running the chains, i.e.~when plotting posteriors and quoting the figure of bias and figure of merit, defined below. 

With regards to the counterterms, we assign uniform $\mathcal{U}(-10,10)$ priors. This choice has no impact on the final results, since the assumed volume $V_\text{fit}$ is large enough to constrain all counterterms of the two-loop model without leading to prior dependence or marginalization effects (see Appendix~\ref{sec:fullMCMC}). 

We run all chains with a maximum of $5\times 10^4$ and $10^6$ steps for the one-loop and two-loop case, respectively, with the affine invariance sampler implemented in \texttt{emcee} \cite{emcee}, using 32 walkers. We also assessed the convergence of the chains by running multiple independent instances and comparing the results, finding negligible differences in posterior means and covariances. 

The figure of bias (FoB) metric is defined as
\begin{eqnarray} \label{eq:FoB}
    \text{FoB}(k_\text{max}) = \bigg( \sum_{\alpha,\beta}(\theta^\alpha - \theta^\alpha_\text{fid})S_{\alpha \beta}^{-1}(\theta^\beta - \theta^\beta_\text{fid})\bigg)^{1/2}\,,
\end{eqnarray}
where $\theta^\alpha_\text{fid}$ are the fiducial (true) values of the simulation parameters and $S_{\alpha \beta}$ is the posterior covariance, for a fixed $k_\text{max}$. Given that observed shifts at the $V_\text{fit}$ volume must arise due to modelling systematics, we adopt a more stringent criterion of 
\begin{eqnarray}\label{eq:fobcrit}
    \text{FoB}(k_\text{max}) < 1\,,
\end{eqnarray}
to consider a model fit acceptable (noting that the $68 \%$ threshold would be $\text{FoB}<1.88$, see e.g. \cite{egge_nonp}).  
We also analyze the figure of merit (FoM) defined as
\begin{eqnarray}\label{eq:fom}
    \text{FoM}(k_\text{max}) = \frac{1}{\big[\text{det}(S_{\alpha \beta}/\theta^\alpha_\text{fid} \theta^\beta_\text{fid})\big]^{1/2}}\,,
\end{eqnarray}
The FoB and FoM are calculated over $\Omega_m,\, H_0$ and $A_s \times 10^9$. 
We also quote the reduced chi-squared statistic
\begin{eqnarray} \label{eq:chi2red}
    \chi^2_\text{red}(k_\text{max}) = \frac{-2\log \mathcal{L}(k_\text{max})}{N_\text{bin}-N_\text{dof}}\,,
\end{eqnarray}
with $N_\text{dof}$ the total number of degrees of freedom and $N_\text{bin}$ the number of bins included. We fix $N_\text{dof} = 6$ for both one and two-loop analysis, corresponding to the five free cosmological parameters and the one-loop counterterm.\footnote{The two-loop counterterms have no impact in the fit at low $k$, where $N_\text{bin}$ is small, and only start being relevant at high $k$, when $N_\text{bin}\gg N_\text{dof}$. We do find that using e.g. $N_\text{dof} = 9$ yields somewhat low values of reduced chi-squared at low $k_\text{max}$, suggesting (incorrectly) that the data is being overfitted at low values of $k_\text{max}$. For a more general discussion, see e.g. \cite{raveri_hu}}. 
We use the \textit{full} simulation volume $V_\text{full}$ in the covariance when computing $\chi_\text{red}^2$, since then the covariance represents the statistical scatter in the power spectrum measurement and in that case $\chi_\text{red}^2$ should be close to unity. 

\subsection{Results for the comparison to simulations} \label{sec:results}

We start by discussing the $k_{\rm max}$ choice for both the one and two-loop modelling of the power spectrum. It is important to emphasize that the choice of $k_{\rm max}$ at a given perturbative order is not independent of the data considered: if the experimental error bars shrink, the range of applicability of the theory at a fixed order is more restricted. As commented, we fix $V_\text{fit}  = 25  \,(\text{Gpc}/h)^3$, roughly resembling the expected DESI volume (sticking to the $z=0$ simulation though). 

\begin{figure}
    \centering
    \includegraphics[width=0.95\linewidth]{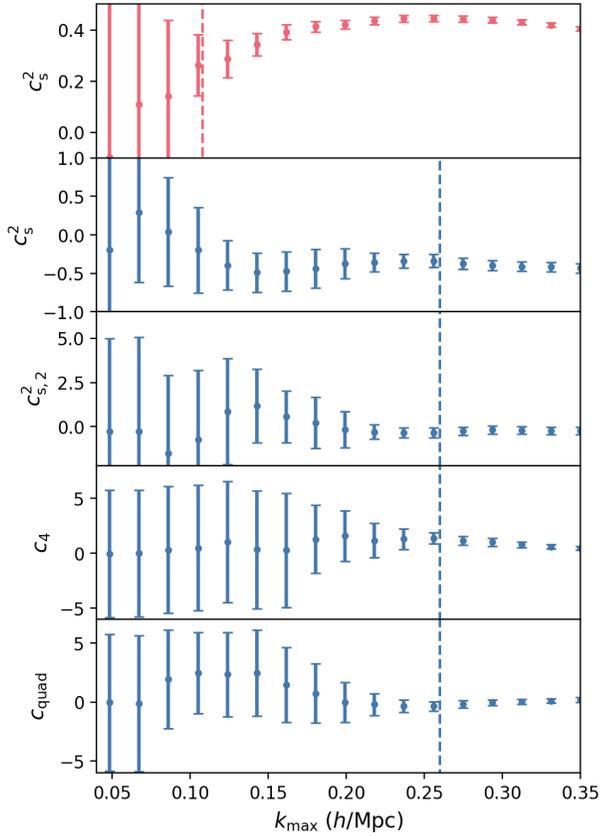}
    \caption{Running of the one and two-loop counterterms as a function of $k_{\rm max}$. The first two panels show the $\propto k^2 P_L(k)$ counterterm (which we label as $c_\text{s}^2$), which is $c_{\text{s},1}^2$ for the one-loop case [in red, cf.~Eq.~\eqref{eq:oneloop}] and $c_{\text{s},1}^2 +c_{\text{s},2}^2$  for the two-loop case [in blue, cf.~Eq.~\eqref{eq:twoloops}]. The remaining panels show the other two-loop counterterms. The vertical lines at $k_\text{max} = 0.11\,h/\text{Mpc}$ and $k_\text{max} = 0.26\,h/\text{Mpc}$ indicate our fiducial choice of maximum wavenumber for the one-loop and two-loop case (see text). For visual purposes, points are slightly displaced horizontally.}
    \label{fig:running_ct}
\end{figure}

One important criteria introduced by \cite{Foreman:2015lca} to choose $k_{\rm max}$ is the running of the counterterms. A strong dependence on $k_{\rm max}$ makes evident the failure of the theory and the need to introduce higher-order corrections \cite{Rubira:2023vzw,Bakx:2025cvu}.  We display in Fig.~\ref{fig:running_ct} the running of the one and two-loop counterterms as fitted in the Dark Sky simulation. For the one-loop case, the single counterterm presents a strong dependence on the running starting at $k_\text{max} > 0.13\,h/\text{Mpc}$. For the two-loop case, the error bars are stable and consistent up to $k_\text{max} < 0.26\,h/\text{Mpc}$.\footnote{Note that the $c_\text{s}^2$ counterterm in the two-loop case is less well-determined at a given wavenumber due to degeneracies with other counterterms. Moreover, its value does not match that of the analogous one-loop counterterm, because we did not subtract the corresponding UV-sensitive part from $P_\text{SPT,2l}$.} For larger values of $k_\text{max}$, we observe a mild running of the counterterms, notably $c_4$ and $c_\text{quad}$. 

\begin{figure}
    \centering
    \includegraphics[width=0.95\linewidth]{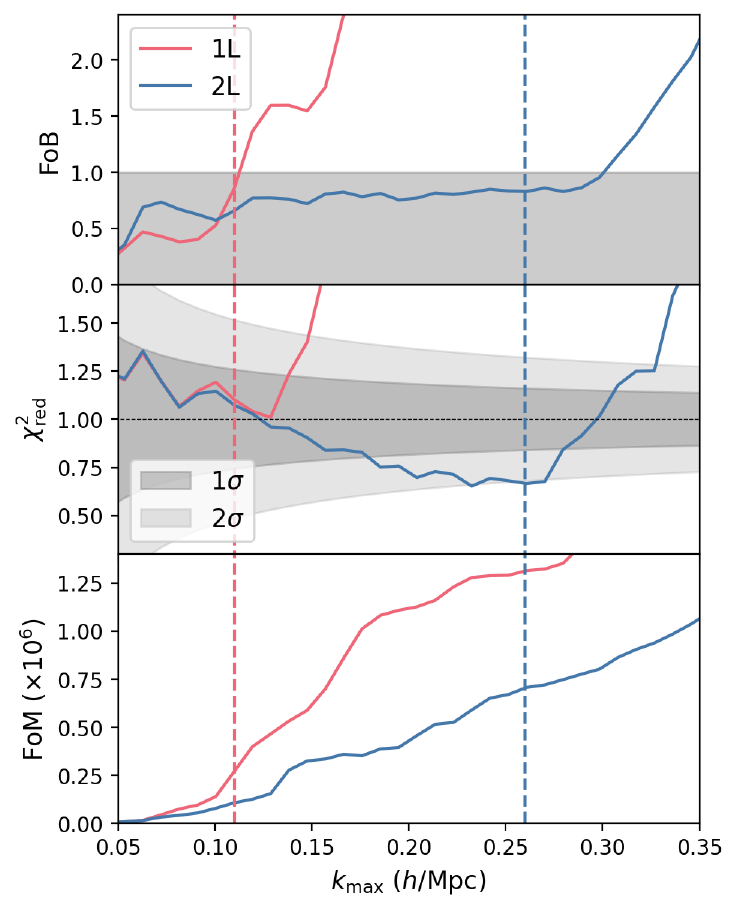}
    \caption{\textit{Top panel}: the FoB defined in Eq.~\eqref{eq:FoB} for the one-loop (red) and two-loop (blue) EFT models, where the shaded band indicates the region satisfying Eq.~\eqref{eq:fobcrit}. \textit{Second panel:} the $\chi^2_\text{red}$ from Eq.~\eqref{eq:chi2red} statistic (computed with the full volume covariance), where the shaded regions indicate the $1\sigma$ and $2\sigma$ confidence regions for the volume $V_\text{full}$. \textit{Third panel:} FoM from Eq. \eqref{eq:fom} as a function of the maximum wavenumber.} 
    \label{fig:residues}
\end{figure}

We also consider two other important criteria for determining $k_{\rm max}$: the FoB statistic and the $\chi^2_\text{red}$, both shown in the top and second panels of Fig.~\ref{fig:residues}, respectively. For the one-loop case, FoB indicates the failure of the one-loop at $k_\text{max} \approx 0.11\,h/\text{Mpc}$, while for the two-loop case, the FoB is essentially constant up to $k_\text{max} < 0.30\,h/\text{Mpc}$ and then increases beyond the FoB $>1$ criteria of Eq.~\eqref{eq:fobcrit} (grey shaded region). We note that for both cases, the FoB sharply increases, indicating an abrupt tendency for shift in the theory parameter space. That is confirmed by the running of the counterterms, discussed above. 
The reduced chi-squared statistics, considered in the second panel, also confirms this sharp failure of the one-loop at $k_\text{max} \approx 0.13\,h/\text{Mpc}$. The two-loop $\chi^2_\text{red}$ starts to sharply increase at $k_\text{max} \approx 0.28\,h/\text{Mpc}$, being within the $2\sigma$ region up to $k_\text{max} < 0.33\,h/\text{Mpc}$. The light and dark gray bands represent the $1\sigma$ and $2\sigma$ region for a volume $V_\text{full}$, respectively.

When determining a value of $k_\text{max}$, one option would be to adopt a rigorous criterion (see e.g. \cite{egge_nonp}) consisting of a combination of all three of the above criteria. However, in practice, some degree of ambiguity is always present. Hence, we simply choose a baseline scale cut that satisfies all three criteria and briefly elaborate on possible other choices at the end of this Subsection.  
As a baseline, we adopt 
\begin{equation}
    k_\text{max} = 0.11\,h/\text{Mpc}\,,
\end{equation}
for the one-loop EFT and 
\begin{equation}    
k_\text{max} = 0.26\,h/\text{Mpc}\,,
\end{equation}
for the two-loop EFT, which correspond in each case to the most conservative values obtained among the running of the counterterms, and the failure of FoB or $\chi^2_\text{red}$. Those values are indicated as vertical dashed lines in Figs.~\ref{fig:running_ct} and \ref{fig:residues}.

Finally, we show in the third panel of Fig.~\ref{fig:residues} the FoM as a function of $k_{\rm max}$. 
The increasing FoM serves as an indication on the information gain in the cosmological parameters as we push the theory limit towards smaller scales. 
The FoM for the one-loop model is evidently larger than the two-loop one at a given $k_\text{max}$, due to the smaller number of nuisance parameters. 
In addition, the one-loop FoM increases faster than the two-loop FoM after our $k_\text{max}$ choice, suggesting the first would benefit most from increasing $k_\text{max}$.
We note however that more aggressive $k_{\rm max}$ choices are discouraged by the sharp deterioration of the FoB and $\chi^2_\text{red}$, with the risk of getting biased in the parameter estimation. 

Note that our range of validity is somewhat smaller than that of \cite{Foreman:2015lca}, who analyzed the same simulation with a two-loop EFT model while fixing the cosmological parameters to the ground truth and found $k_\text{max} = 0.34\,h/\text{Mpc}$ based on the running of their counterterms being stable up to that scale. 
When fixing cosmological parameters to the ground truth inferring only the counterterms and implementing the Lagrangian IR-resummation procedure used in \cite{Foreman:2015lca}, we find reasonable agreement with their result. 
We further note that the two-loop FoB is below unity up to $k_\text{max} < 0.31\,h/\text{Mpc}$, and $\chi^2_\text{red}$ is still within the $2\sigma$ agreement up to $k_\text{max} < 0.33\,h/\text{Mpc}$ indicating reasonable agreement between our results and \cite{Foreman:2015lca}. 
Nevertheless, we have attempted to be conservative with our choices of scale cuts to prevent overfitting. 

\begin{figure}
    \centering
    \includegraphics[width=\linewidth]{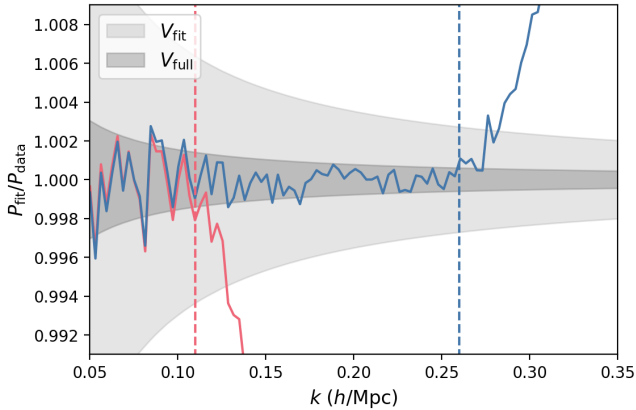}
    \caption{Ratio between the fitted power spectra and the simulation data at the fiducial scale cut. The shaded regions indicate the $1\sigma$ error for $V_\text{fit}  = 25  \,(\text{Gpc}/h)^3$ and for the full simulation box. The red and blue vertical dashed lines indicate the fiducial scale cut choices (see text).}
    \label{fig:ratiofit}
\end{figure}

We also consider in Fig.~\ref{fig:ratiofit} the ratio between the best-fit model of the one- and two-loop spectra from maximizing the log-likelihood and the simulation data for the $k_\text{max}$ choices indicated by dashed vertical lines. The light and dark gray bands represent the $1\sigma$ region for $V_\text{fit}$ and $V_\text{full}$, respectively. In accordance with the $\chi_\text{red}^2$, we see that the ratio of the simulation data and the model is very compatible with unity, being within the error bars chosen up to their respective $k_\text{max}$ values. We note the achievement of modeling the power spectra at per-mille level.

\begin{figure}
    \centering
    \includegraphics[width=0.95\linewidth]{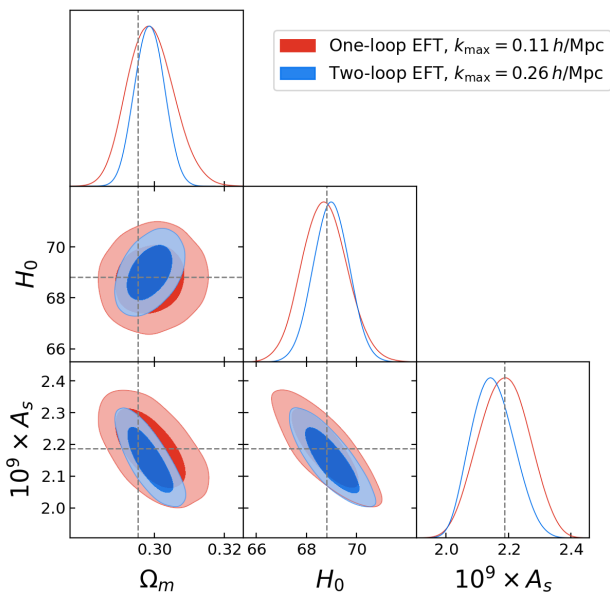}
    \caption{A comparison of the constraining power of the one-loop and two-loop EFT models for the power spectrum on their maximum range of validity, namely $k_\text{max} = 0.11 \,h/\text{Mpc}$ and $k_\text{max} = 0.26 \,h/\text{Mpc}$, respectively. Shaded regions indicate the $1$ and $2\sigma$ confidence regions.}
    \label{fig:mcmc}
\end{figure}

We show in Fig.~\ref{fig:mcmc} the resulting parameter posteriors on $\Omega_m,H_0$ and $A_s$ for the baseline scale cut choice (see App.~\ref{sec:fullMCMC} for the full posteriors). The improvement of the two-loop analysis is visually apparent; we are able to access smaller scales and, despite marginalizing over four more counterterms, extract more cosmological information. We find for $\Omega_m,H_0$ and $A_s$ an improvement of approximately $35\%, 20\%$ and $15\%$, respectively, compared to the one-loop, providing unbiased constraints in all parameters. 
The FoM at $k_\text{max} = 0.26\,h/\text{Mpc}$ for the two-loop case is larger by a factor of $\sim 2.6$ than the one-loop FoM at $k_\text{max} = 0.11\,h/\text{Mpc}$. If one relaxes the FoB requirement from Eq. \eqref{eq:fobcrit} to extend the one-loop reach to $k_\text{max} = 0.12\,h/\text{Mpc}$ and push the two-loop analysis to correspondingly shorter scales $k_\text{max} = 0.31\,h/\text{Mpc}$ (thus matching the one-loop FoB, and ignoring the mild $\sim 1\sigma$ drifts in the two-loop counterterms), the improvement in the FoM mildly degrades to a factor of $\sim 2.2$.  

\section{Discussion and conclusion}\label{sec:disc}
In this work we have extended the \texttt{COBRA} framework to calculate the two-loop power spectrum. Our pipeline computes a single two-loop power spectrum in $\sim 1$ ms and yields converged posteriors in a negligible $\sim 20$ hours on one core. 
For that, we precomputed $O(100)$ two-loop integrals for the SVD basis, each of those taking a few minutes per mode. Thus, we conclude that a calculation of the two-loop power spectrum of biased tracers is also within computational reach. 
We have demonstrated that \texttt{COBRA} is both fast and accurate, enabling the extension of EFT of LSS application using real data (so far limited to one-loop) to the two-loop case in the near future.

In addition, we have shown that additional unbiased cosmological information can be extracted from the matter power spectrum by including two-loop corrections. Adding two-loop corrections shrink posteriors on $\Lambda$CDM parameters by $15-35\%$ and increase the FoM (calculated for $\Omega_m,H_0 $ and $A_s$) in between $2.2$ and $2.6$. Therefore, this work adds important piece to answer how much information can be extracted from the mild nonlinear using galaxy clustering data in real space. 

Another relevant ongoing question is to estimate the range of applicability $k_{\rm max}$ of perturbative approaches. We find a sharp deterioration of the one-loop FoB for $k_\text{max} > 0.11\,h/\text{Mpc}$ for volumes comparable to DESI at $z=0$. For two loops the deterioration of FoB and $\chi^2_{\rm red}$ is less severe, but the running of the counterterms led us to adopt a conservative $k_\text{max} = 0.26\,h/\text{Mpc}$. Therefore, the two-loop result more than doubles the range of validity of the EFT of LSS. Doubling the scales of applicability of EFT modeling not only has the advantage of decreasing the error bars of $\L$CDM parameters, but also allows for testing models that affect the power spectrum on quasi-linear scales, e.g.~by probing the suppression or enhancement of matter fluctuations. This is especially of interest for currently ongoing efforts to measure CMB lensing (see e.g. \cite{qu_lensing,madh_lensing,SPT-3G:2024atg}) and galaxy lensing (e.g.~\cite{chen_lensing}) on mildly non-linear scales. 

Despite the added constraining power associated with the larger range of scales probed by the two-loop matter power spectrum (third panel of Fig.~\ref{fig:residues}), most of this increase is reflected in tighter errors on the EFT counterterms rather than the FoM on cosmological parameters of interest. This is of course partly expected, since the counterterms have most of their support at larger wavenumbers. For example, one can clearly see in Fig. \ref{fig:running_ct} (bottom panel) that the posterior width on $c_\text{quad}$ shrinks by a factor of few between $k_\text{max} = 0.12\,h/\text{Mpc}$ and $k_\text{max} = 0.26\,h/\text{Mpc}$. Thus, it would be valuable to search for models that contain fewer nuisance parameters than the EFT counterterm parametrization, either with the Lagrangian framework that incorporates shell crossing effects (e.g. \cite{McDonald:2017}) or alternatively with the recently proposed framework of Vlasov Perturbation Theory (VPT) \cite{garny2025vlasov, vpt_lin}, either being a natural candidate for such models. 

It should be emphasized that the results presented in this work cannot readily be generalized to draw conclusions about the two-loop galaxy power spectrum, which involves many more nuisance parameters, both for the one- and two-loop results (see e.g. \cite{Eggemeier:2018qae, Schmidt:2020tao,donath,Bakx:2025cvu}). 
In addition, the presence of redshift-space distortions may limit the range of validity of perturbative models compared to the real-space case considered here \cite{Okumura:2012, Vlah:2013,damico_rsd,ivanov_fog, Chen:2020}, at the same time adding information about the growth rate. 

\begin{acknowledgments}
We thank Mathias Garny and Constantinos Skordis for interesting discussions during the early stages of this work. This publication is part of the project ``A rising tide: Galaxy intrinsic alignments as a new probe of cosmology and galaxy evolution'' (with project number VI.Vidi.203.011) of the Talent programme Vidi which is (partly) financed by the Dutch Research Council (NWO). For the purpose of open access, a CC BY public copyright license is applied to any Author Accepted Manuscript version arising from this submission. H.R.~and Z.V.~acknowledge the support of the Kavli Foundation.
\end{acknowledgments}

\appendix

\section{Constructing loop tensors}\label{sec:tensors}
Here we provide more details on how loop tensors of any rank can be efficiently constructed with \texttt{COBRA}. 
Note first that the tensor $\mathcal{S}_{i_1i_2\dots i_k}$, constructed as a linear functional of direct tensor products of $v_{i_1},\dots,v_{i_k}$, can always be taken to be symmetric.
A symmetric tensor of rank $k$ in $N$ dimensions has 
\begin{eqnarray}\label{eq:tensdim}
    \binom{N+k-1}{k}
\end{eqnarray}
independent elements; these can be taken to be the `super-diagonal' ones, i.e. $\mathcal{S}_{i_1i_2\dots i_k}$ with $i_1 \geq i_2 \geq \dots \geq i_k$. 
For the two-loop tensors ($k=3$), the evaluation with $N=9$ elements in the basis for the full spectra $N=6$ for the no-wiggle demands then the computation of 165 and 56 integrals, respectively for every $k$. 
At first sight, to compute all entries of this tensor, one would seemingly need to explicitly compute all `cross-correlations' of the form 
\begin{eqnarray}
    \mathcal{S}_{i_1i_2\dots i_k} = \mathcal{S}[v_{i_1}, v_{i_2}, \dots ,v_{i_k}] \quad (i_1 \geq i_2 \geq \dots).
\end{eqnarray}
where multiple input scale functions $v_i$ are used. 
This is technically and numerically demanding, especially if one breaks the IR safety of integrands (see Sec.~\ref{app:IRsafe} for further discussion).
However, it is possible to construct the full tensor without requiring multiple input scale functions, as we discuss next. 

The procedure we outline here works for tensors of any rank, but we only explicitly need it for ranks 2 and 3.
To see how this works, note that $\mathcal{S}$ is always linear in each of its inputs. For that reason, we can construct `off-diagonal' elements of the symmetric tensor $\mathcal{S}$ (i.e. where the indices are not all equal) by taking linear combinations of $\mathcal{S}$ evaluated on the \textit{same} element. As an example, one has for rank 2:
\begin{eqnarray}\label{eq:constr}
    2\, \mathcal{S}_{ij} &=& \mathcal{S}_{ij} + \mathcal{S}_{ji} = \mathcal{S}[v_i,v_j] + \mathcal{S}[v_j, v_i]\\ \nonumber
    &=& \mathcal{S}[v_i+v_j,v_i+v_j]  
    -\mathcal{S}[v_i,v_i] - \mathcal{S}[v_j,v_j]\,,
\end{eqnarray}
where the bottom line only involves evaluations of $\mathcal{S}$ with the same element in each position. Hence, by evaluating $\mathcal{S}$ for all possible linear combinations $v_i + v_j$ with $i \geq j$ as both first and second argument \footnote{When $i=j$ we can recover $\mathcal{S}[v_i,v_i] = \frac{1}{4}\mathcal{S}[v_i+v_i,v_i+v_i]$.} [which reproduces the correct dimension counting from Eq.~\eqref{eq:tensdim}] and then applying Eq.~\eqref{eq:constr}, we can automatically obtain all entries $\mathcal{S}_{ij}$.

In general, one can prove (see e.g.~\cite{Marcus1973}) that evaluating the tensor $\mathcal{S}_{i_1 i_2 \dots i_k}$ amounts to evaluating $\mathcal{S}$ for the linear combinations 
\begin{eqnarray}
    v_{i_1} + v_{i_2} + \dots + v_{i_k} \quad (i_1 \geq i_2 \geq \dots \geq i_k)\,,
\end{eqnarray}
as each of its arguments. Indeed, the number of such combinations equals the expression in Eq.~\eqref{eq:tensdim}. Thus, no fewer combinations can be used, and using more is redundant. Then, by taking appropriate linear combinations of these elements one can recover the elements $\mathcal{S}_{i_1 i_2 \dots i_k}$. 
\begin{figure*}[t]
    \centering
    \includegraphics[width=\linewidth]{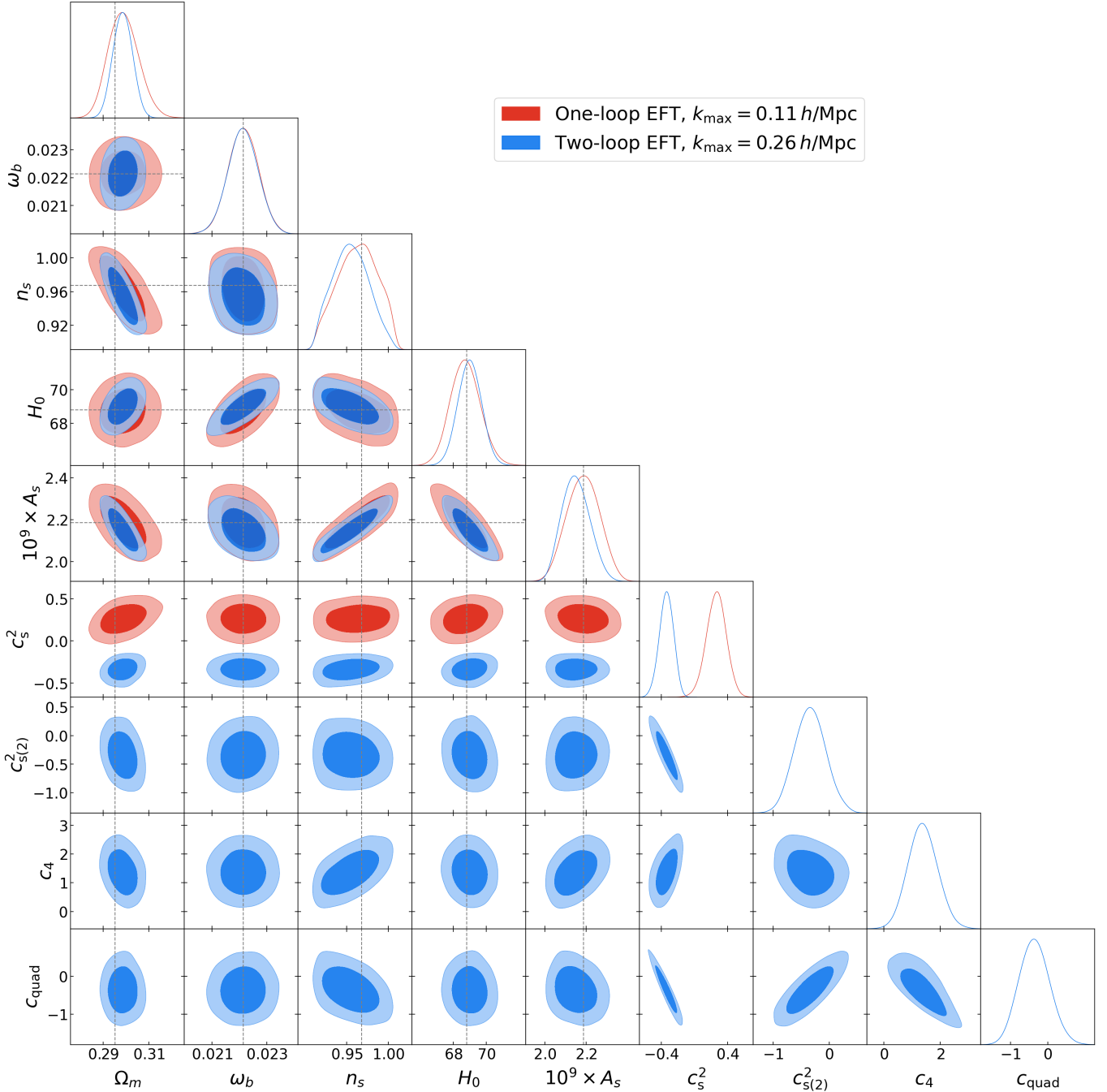}
    \caption{Full posteriors for the one-loop (red) two-loop (blue) MCMC at their maximum wavenumbers. 
    }
    \label{fig:fullpost}
\end{figure*}

\section{Infrared safety} \label{app:IRsafe}
It is well known that cosmological perturbation theory, both in its SPT and EFT formulations, exhibits so-called infrared (IR) divergences \cite{Jain:1995, Carrasco:2013} at the level of individual loop diagrams. These arise from large-scale displacements that cancel out in equal-time correlators at each order in perturbation theory, due to the equivalence principle. In practice, this leads to partial cancellations of power in the high-$k$ regime of the power spectrum when summing over all relevant loop diagrams. As emphasised in \cite{Carrasco:2013} (see also \cite{Fasiello:2022}), it is preferable to implement these cancellations directly at the level of the integrands during the evaluation of the power spectrum with fixed accuracy, rather than relying on the cancellation of individual diagram contributions, which can lead to a loss of numerical precision.

Consequently, one might ask whether introducing the \texttt{COBRA} decomposition given in Eq.~\eqref{eq:decomp} could exacerbate these IR divergences. We show that this is generally not the case and that the IR cancellation persists for each symmetrized cross-contribution of the two different basis functions.

We start with the one-loop result. IR divergent contribution of the $P_{22}$ diagram comes from two integration regions 
\begin{align}
P_{22}(k) \subset 2 \int_{\bm p} \left[ F_{2}(\bm p, \bm k - \bm p) \right]^2 v_i(\bm p) v_j(\bm k - \bm p) 
= \int_{p < |\bm k - \bm p|} [\,\dots]+ \int_{p \geq |\bm k - \bm p|}[\,\dots] \, , 
\end{align}
which, as $p\to 0$, give the two asymptotic behaviors, one proportional to $(k^2/p^2)v_i(k)v_j(p)$ and the other to $(k^2/p^2) v_i(p)v_j(k)$. 
Note that symmetrizing this contribution in the $(i,j)$ indices yields an equivalent result.
The two contributions above are exactly cancelled by the $P_{13}$ diagrams, where we have 
\begin{align}
P_{13} + P_{31} &\subset 3 \int F_{3}(\bm k, \bm p, - \bm p) \Big( v_i(k)v_j(p) + v_i(p)v_j(k) \Big) \, .
\end{align}
Since this cancellation occurs for each $(i,j)$ pair, it also holds for the full sum. Crucially, the cancellation must be preserved at the level of each pair, not only after summing over the coefficients $w_i$ in, e.g. Eq.~\eqref{eq:1loop1}. This ensures there are no large cancellations between various terms in the $S^{q}_{ij}$ tensor and allows the same numerical setup used for the full $P_L^\Theta$ to be applied when precomputing the components of the $S^{q}_{ij}$ tensor. This property can also be made manifest in the resulting integrals, by using the multilinearity of the $S_{ij}$ tensor as described in Sec.~\ref{sec:tensors}. By expressing the cross-terms of the $S_{ij}$ tensor as a linear combination of several auto-correlations we obtain the form that is explicitly IR safe in each term.

Similar analyses can be carried out for the two-loop power spectrum. As discussed in refs. \cite{Carrasco:2013, Fasiello:2022}, the IR cancellation occurs separately between the part of the $P_{24}$ and $P_{33,{\rm II}}$ contributions, as well as between the remaining parts of $P_{24}$, $P_{33,{\rm II}}$, and $P_{15}$. The only difference is that, when remapping the integration regions, some IR-divergent regions cannot be combined and must be treated separately, because exchanging $v_i$ and $v_j$ does not yield the same contribution. This feature already appears in the one-loop result, as shown explicitly above. Importantly, the cancellation again occurs at the level of each $(i,j,k)$ triplet contributing to the sum in the last term of Eq.~\eqref{eq:2loop1}. This ensures that, for fixed accuracy, the contributions to the $\mathcal{S}^{c}_{ijk}$ tensor can be precomputed using the same numerical setup as for the original linear power spectrum $P_L^\Theta$. Moreover, as was also the case for one-loop, the multilinearity of the $S_{ijk}$ tensor can again be employed to recast the cross-terms as a linear combination of auto-correlations as shown in Sec.~\ref{sec:tensors}. This again renders the result manifestly IR-safe.

\section{Full posteriors}\label{sec:fullMCMC} \label{app:fullMCMC}

Here we display the full posteriors of the one-loop and two-loop MCMC in Fig. \ref{fig:fullpost}. It is evident that the data is able to constrain all nuisance parameters beyond their prior ranges. We emphasize again that we do not expect the $c_\text{s}^2$ parameters to match between the one-loop and two-loop fits. We also see that at this precision, $n_s$ and $\omega_b$ cannot be constrained meaningfully, although the posterior for $n_s$ is markedly different from its (uniform) prior in Eq. \eqref{eq:nsprior}. Note that for the two-loop model, the counterterms (lower right 4 by 4 triangle) are clearly correlated among each other, as are the cosmological parameters (upper left 5 by 5 triangle), but at least in the 2D marginalized posteriors the counterterms and cosmological parameters are mostly uncorrelated (lower left 4 by 5 rectangle), with the exception of e.g. $A_s$ and $c_4$ or $n_s$ and $c_4$. 


\bibliographystyle{JHEP}
\newpage%
\bibliography{mybibliography}
\end{document}